\documentclass[nanomaterials,article,submit,moreauthors,pdftex]{Definitions/mdpi}
\firstpage{1}
\pubvolume{0}
\issuenum{0}
\articlenumber{0}
\doinum{0}
\pubyear{2021}
\copyrightyear{2021}
\updates{yes}
\usepackage{changes}
\usepackage{upgreek}

\Title{Preparation and fluorescent wavelength control of multi-color Nitrogen-doped carbon nano-dots}

\Author{Wenli Li$^{1,\dag}$,  Ju Tang$^{1,2,\dag}$, Yuzhao Li$^1$,
Han Bai$^1$, Weizuo Zhang$^1$, Jin Zhang$^{1,3,}$*,
Yiming Xiao$^{1,}$*\orcidA{}, and Wen Xu $^{1,4,5,}$*}

\AuthorNames{Wenli Li, Ju Tang, Yuzhao Li, Han Bai, Weizuo Zhang,
Jin Zhang, Yiming Xiao, and Wen Xu}

\address{
$^{1}$ \quad School of Physics and Astronomy, Yunnan University, Kunming 650091, China;
18388308185@163.com (W.L.);  jutang\_ynu@163.com (J.T.);
liyuzhao@mail.ynu.edu.cn (Y.L.); bh001925@163.com (H.B.); zhangwz@ynu.edu.cn (W.Z.) \\
$^{2}$ \quad Department of Physics, School of Electrical and Information
Technology, Yunnan Minzu University, Kunming 650504, China\\
$^{3}$ \quad Yunnan Carbon Base Technology Co. LTD, Kunming 650028, China \\
$^{4}$ \quad Micro Optical Instruments Inc., 518118 Shenzhen, China \\
$^{5}$ \quad Key Laboratory of Materials Physics, Institute of Solid State
Physics, HFIPS, Chinese Academy of Sciences, Hefei 230031, China}

\corres{Correspondence: zhangjin@ynu.edu.cn; yiming.xiao@ynu.edu.cn; wenxu\_issp@aliyun.com}
\firstnote{These authors contributed equally to this work.}

\abstract{It is known that, by taking advantage of heteroatom doping, the
electronic states and transition channels in carbon nano-dots (CNDs) can
be effectively modulated. Thus, the photoluminescence (PL) properties of
CNDs can be changed. For potential applications of CNDs as advanced materials
for optoelectronic devices, it is important and significant to develop the
practical techniques for doping heteroatoms into CNDs. In this work, we
synthesize the multi-color fluorescent by using a fast and effective microwave
method where the CNDs are nitrogen-doped. We examine the influence of different
ratios of the raw materials on the structure and optical properties of N-CNDs.
The results show that the prepared N-CNDs can generate blue (445 nm),
green (546 nm), and orange (617 nm) fluorescence or PL with the
mass ratio of the raw materials at 1:1, 1:2 and 1:3, respectively. We
find that the N content in N-CNDs leads to different surface/edge states
in $n-\pi^*$ domain. Thus, the wavelength of the PL emission from N-CNDs
can be tuned via controlling the N concentration doped into the CNDs.
Moreover, it is shown that the intensity of the PL from N-CNDs is mainly
positively related to the content of C-O groups attached on the surface/edges
of the N-CNDs. This study provides an effective experimental method and
technical way to improve the fluorescent emission, and to modulate the color
of the PL emission from CNDs.}

\nolinenumbers
\begin{document}
\section{Introduction}

Carbon nano-dots (CNDs) are new types of carbon-based
nanomaterials \cite{Xu04,Li12}. The structure of CNDs is generally considered to be composed of
$sp_2$/$sp_3$ carbon and oxygen/nitrogen based-groups or polymer groups \cite{Zhu15}.
This means that CNDs have better practical applications in many fields, such as light-emitting
devices, biological imaging, photothermal therapy, photocatalysis, electrochemical
energy storage, biomedical \cite{Sun06} and visual
precision pH sensing \cite{Zhao21}. In particular, due to the flexibility of
surface modification and the tunable PL emission wavelength, CNDs have gradually become
the most suitable alternative material for the application of metal nanoclusters and
traditional dye molecules in the field of optoelectronics \cite{Liu20}.

Due to the structural differences caused by different raw materials and
experimental conditions, CND systems often have different PL emission behaviors.
In general, the PL properties of CNDs depend strongly on the carbon source, synthesis
technology, parameter setting in the preparation process, and edge/surface
modulation \cite{Wu14,Qian14}. More specifically, the intensity and frequency positions
of PL emission of CNDs are subject to the existence of different functional groups or
edge states, the interaction between the electronic states of $sp_2$ conjugated domain,
the interaction between the electronic states in surface state chemical groups/hanging bonds,
and the properties of fluorophores in CNDs \cite{Li14,Hu20}. However, there is no
final conclusion on the photoluminescence mechanism of CNDs. It is generally believed
that the photoluminescence mechanism of CNDs depends on quantum confinement
effect~\cite{Shen12}, surface state \cite{Zhu11,Pan10,Kozak13,Li13,Liu14,Dong16}, and molecular
state \cite{Wang17}. It is worth noting that the existence of surface/defect states is
the potential reason for the PL excitation dependent luminescence characteristics. The
fluorescence of surface/defect states as PL emission centers can be generated directly
by optical pumping excitation, or by the energy transfer of eigenstates \cite{Bhunia13,Purbia16}.
Therefore, the preparation of tunable fluorescent CNDs and the disclosure of its optical
mechanism are particularly important.

Recently, the effect of the pH values on fluorescence properties of CNDs has also been
investigated \cite{Xiao17,Yogesh17,Zhao21}. Wang et al. \cite{Wang17} reported that the
reflux of a solution containing L-cysteine and galactose with NaOH at different
concentrations can emit PL with different colors. The PL wavelength generated from CNDs
would be shorter with increasing NaOH concentration. By the way, temperature also
plays an important role in carbon precursor carbonization or cutting up
carbon/graphite materials for the preparation of CNDs. Reaction temperature
during the synthesizing of CNDs can affect the chemical, physical and bio properties
of CNDs, such as the degree of carbonization or crystallization, particle size, types
of surface/edge functional groups, the corresponding PL strength, and peak
position~\cite{Bhunia13,Purbia16}. Generally, with increasing reaction temperature, it
can gradually increase the contents of special hetero-atoms or produce a new molecular
structure in the preparation of CNDs \cite{Sun13}. For example, the preparation of
nitrogen-doped CNDs at high temperatures 600--900 $^{\circ}$C can form a variety of
molecular structures in the carbon structure, such as pyridine, pyrrole or graphite
nitrogen \cite{VanKhai12}.
It has been noticed that the synthesis method can also
affect the wavelength of the PL emission from CNDs prepared with the same initial
precursor \cite{Shi16}.
The emission spectra of carbon dots dependent on the excitation wavelength
for most cases and only exhibit excitation-independent behavior when CNDs were fully
surface-passivated \cite{Kundelev19,Miao18}. Typically, the structure of citric acid (CA) contains multiple
hydroxyl and carboxyl groups, and it is one of the most commonly used precursors for carbon dots preparation by a bottom-up method \cite{Kundelev19}. If CA and urea are used as precursors, by varying the mole ratio of CA/urea and the reaction temperature, the carbon dots can emit blue to red light, covering the entire light spectrum. When the
reaction temperature is at 200 $^{\circ}$C and the mole ratio of CA/urea is greater than 0.7, red carbon dots with the maximum emission peak at 630 nm have the fluorescence quantum efficiency of 12.9\%. It suggests that temperature and
mole ratios of CA/urea can tune the maximum emission of carbon dots from blue--green to
red \cite{Miao18}. Inspired by the above research, we propose a simpler and more environmentally friendly method to prepare CNDs with adjustable luminous intensity and wavelength. In this work, we use the citric acid and L-glutamic acid as carbon resources to fabricate the CNDs. By controlling the ratio between citric acid and L-glutamic acid, the multi-color N-CNDs can be obtained in deionized water as a dispersing agent.
From a viewpoint of material application, it is important
and significant to understand and to examine the dependence of the intensity and
wavelength of the PL emission from CNDs on the ratio of the raw materials. This
becomes the prime motivation of the present study.

\section{Experiment}
\label{sec:Experiment}

\subsection{Raw Materials}
The analytically pure citric acid and L-glutamic acid were purchased
from Afisa  Chemical Limited Liability Company (Tianjin, China). The deionized
water was produced in the laboratory.

\subsection{Preparation of N-CNDs by Means of Microwave and Experimental Measurements}

\begin{figure}[H]
\includegraphics[width=0.7\linewidth]{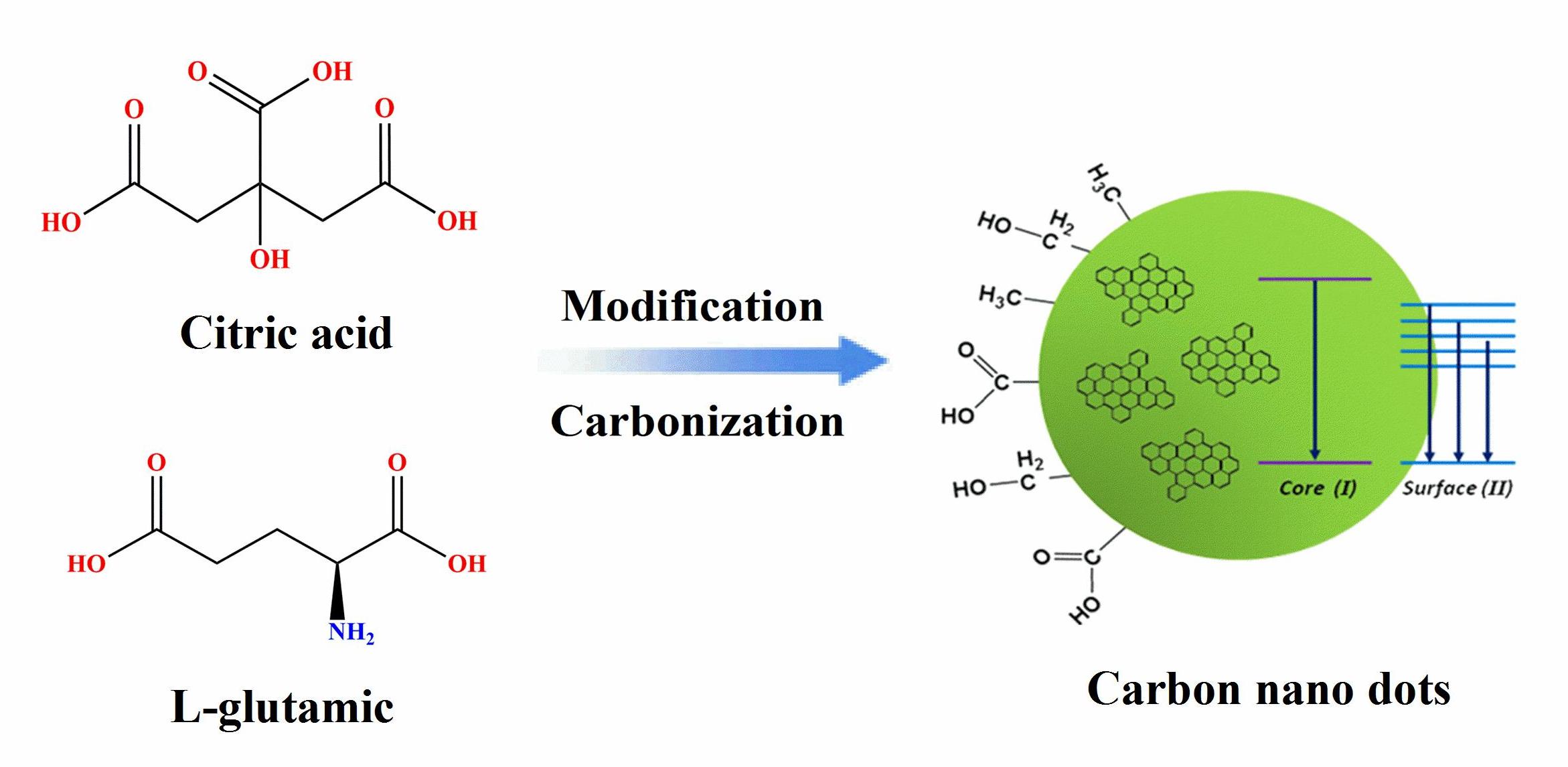}
\caption{
{The diagram of the preparation procedure of CNDs.}}\label{fig1}
\end{figure}

The general processes to synthesize the CNDs from citric acid and
L-glutamic acid are as follows: (i) Citric acid and L-glutamic acid were
mixed evenly with the mass ratio at 1:1 (1 g:1 g), 1:2 (1 g:2 g) and 1:3 (1 g:3 g), respectively.
Here, the mixture is put into the beaker and placed into the microwave for constant heating.
(ii) After being clarified and fully reacted in a microwave reactor for 5 min, the microwave
heating temperature is 300~degrees Celsius, and the power is 800 watts. (iii) We let
the stuff in the beaker cool down naturally till room temperature and add 10 mL deionized
water into the beaker. (iv)~The mixture is magnetically stirred for 10 min to achieve
the uniform and full mix of the matters and water. The mixture is further centrifuged
at a speed of 12,000 r/min for 30 min. After purifying by the dialysis bag
(the aperture
is 30,000 D), we can obtain the solutions which contain CNDs. The diagram of
the preparation procedure of CNDs is shown in Figure \ref{fig1}. We obtain the N-CNDs, which
can emit blue, green and red fluorescent light, respectively.

The PL
spectra of N-CNDs
solutions were measured by a fluorescence spectrometer F9818012
(SHANGHAI LENGGUANG TECHNOLOGY CO., LTD, Shanghai, China).
The X-ray photoelectron spectroscopy (XPS) of N-CNDs was measured by using
PHI5000 Versa Probe II photoelectron spectrometer (Thermo Scientific, New York, NY,  USA) with Al $K_{\alpha}$ at 1486.6 eV.
The morphology and micro-structures of the CNDs were characterized by using
the JEM 2100 transmission electron microscopy (JEOL, Tokyo, Japan), with an accelerating
voltage of 300 kV. The Fourier-transform infrared (FT-IR) spectra of the CNDs
were recorded on a Perkin Elmer TV1900 instrument (Thermo Scientific, New York, NY,  USA).
The UV-Vis absorption spectra were measured by a Specord 200 UV-Vis spectrophotometer
(Germany Jena (Zeiss) Co., Jena, Germany).

\section{Results and Discussions}
\begin{figure}[H]
\includegraphics[width=0.5\linewidth]{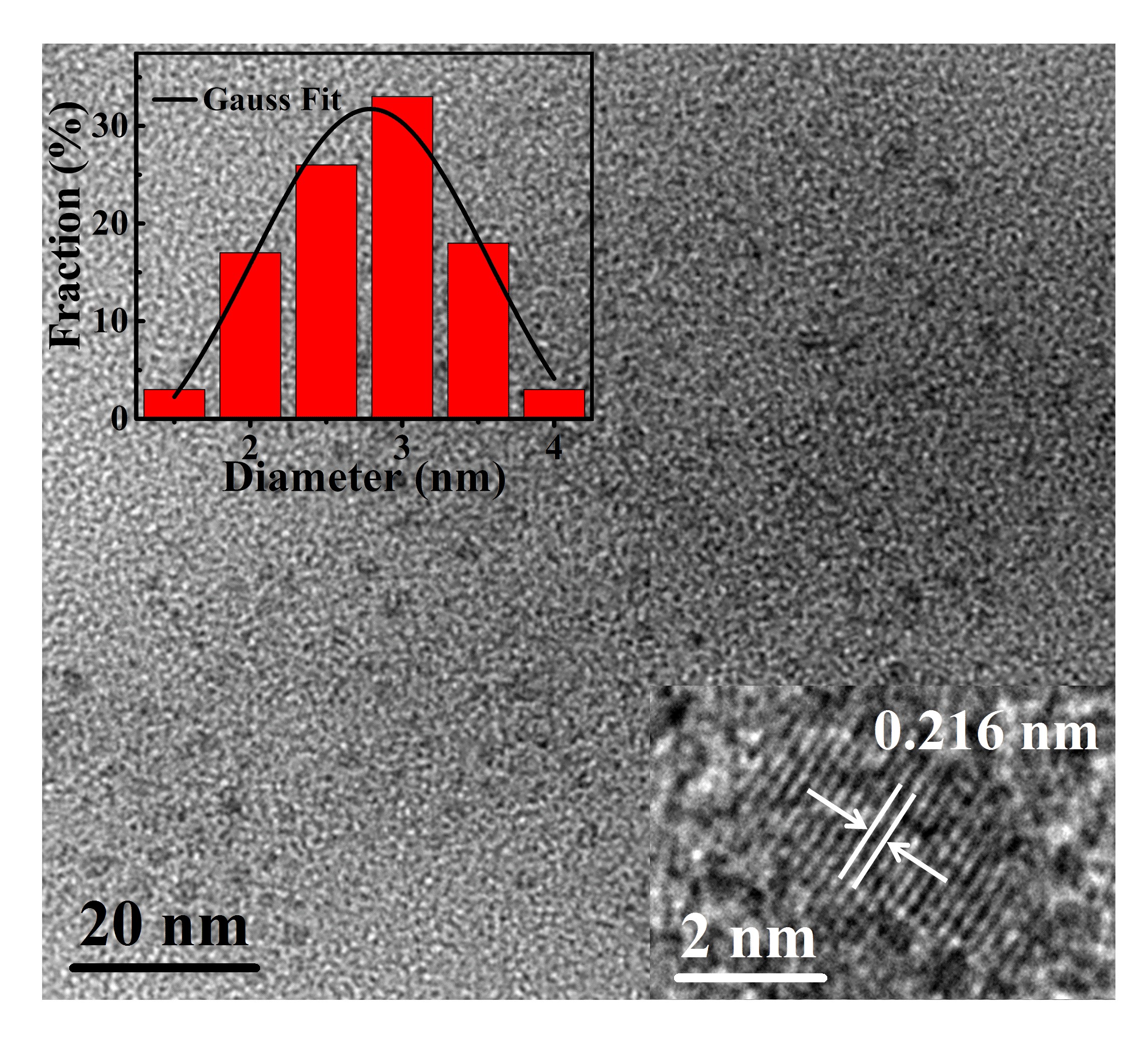}
\caption{TEM images of N-CNDs, their lattice fringes (HRTEM image, lower
inset), and the diameter distribution (upper inset).}\label{fig2}
\end{figure}

In Figure \ref{fig2}, we show the TEM images of morphology and lattice stripes
(the inserted image, HRTEM image) of N-CNDs, and particle size distribution
(the inserted chart) of CNDs. Here,the N-CNDs were fabricated by mixing
citric acid with L-glutamic acid at the ratio of 1:3. We can see that N-CNDs
are with clear crystal structures and their morphology is disk-like. The size of CNDs
is in the range of 1.5--4.0 nm, with an average size of 2.6~nm. From the
HRTEM image of single N-CND (see lower inset), the well-resolved lattice fringes with inter-planar
spacing of 0.216 nm were obtained. This value is close to the (100) diffraction
facets of graphite carbon, which indicates that the carbon core of the N-CNDs is
with good crystallinity \cite{Guo16}.

\begin{figure}[H]
\includegraphics[width=1\linewidth]{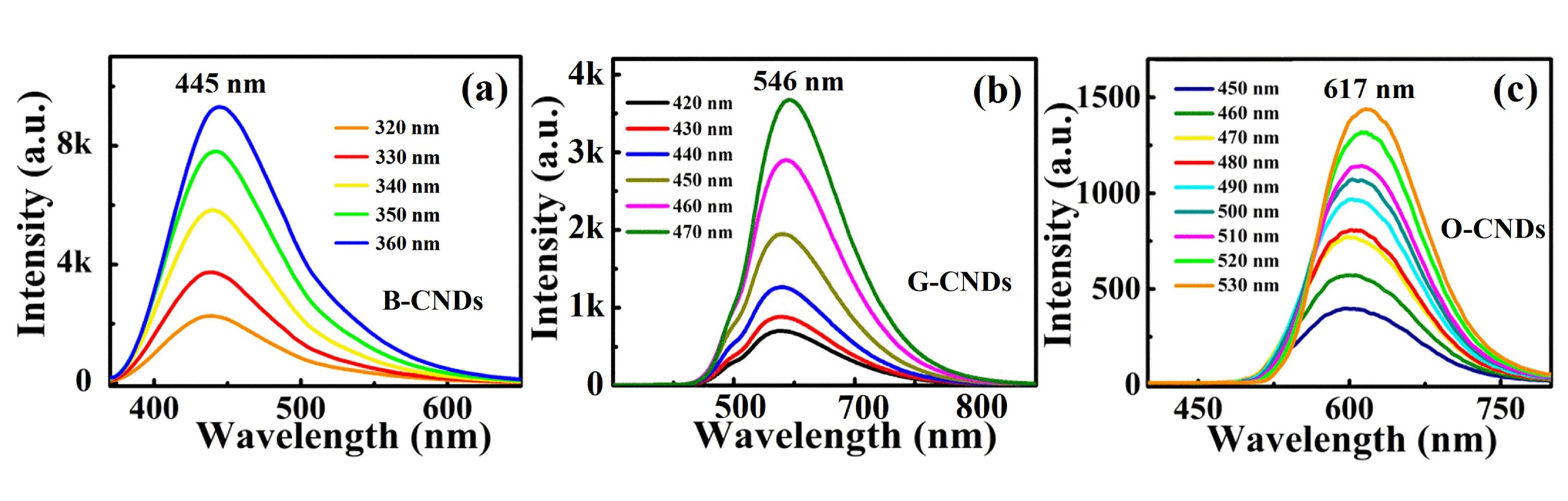}
\caption{The PL spectra of (\textbf{a}) B-CNDs, (\textbf{b}) G-CNDs, and (\textbf{c}) O-CNDs prepared by the mixture of citric acid and L-glutamic acid with the mass ratio of 1:1, 1:2, and 1:3, respectively.}
\label{fig3}
\end{figure}

\begin{figure}[H]
\includegraphics[width=0.6\linewidth]{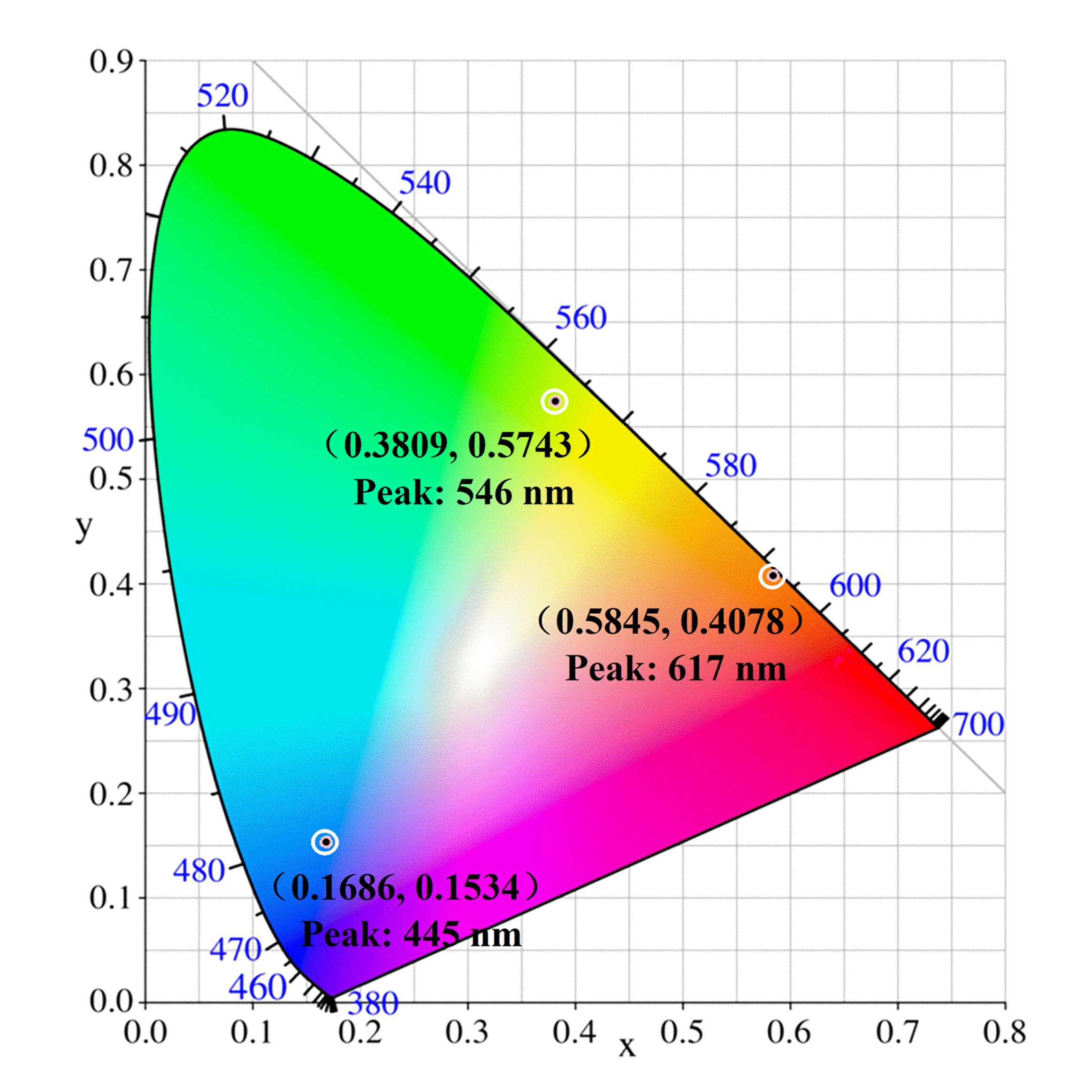}
\caption{\textls[-35]{The chromaticity diagrams of N-CNDs with different
raw ratios at optimal excitation~wavelength.}}
\label{fig4}
\end{figure}

Figure \ref{fig3}a--c shows the fluorescence emission spectra of blue-CNDs
(B-CNDs), green-CNDs (G-CNDs), and orange-CNDs (O-CNDs) prepared by the
mixture of citric acid and L-glutamic acid, with the ratio of 1:1, 1:2 and 1:3,
respectively. When B-CNDs, G-CNDs, and O-CNDs are excited by the wavelengths
of 360 nm, 470 nm and 530 nm, the corresponding peak position of fluorescence
emissions from B-CNDs, G-CNDs, and O-CNDs are located at wavelength of 445 nm,
546 nm, and 617 nm, respectively. The fluorescence intensity of them are
increasing with increasing of excitation wavelength for three types of CNDs.
The peak positions of PL for these CNDs are 
{slightly shift with different} 
excitation wavelengths. 
{During the PL measurement, the influence of UV exposure on the PL intensity of the B-CNDs, G-CNDs, and O-CNDs had also been examined. The PL intensity does not vary under continuous UV irradiation for hours for B-CNDs, G-CNDs, and O-CNDs in deionized water.
This indicates that the CNDs prepared are quite stable and would have good photostability for the practical applications.} The results imply that the fluorescence properties induced by
N impurity in these N-CNDs are quite stable. Furthermore, in \mbox{Figure \ref{fig4}},
we show the chromaticity diagrams of N-CNDs, with different raw ratios at optimal
excitation wavelengths. The color coordinates shown in the chromaticity diagram
indicate that the B-CNDs, G-CNDs, and O-CNDs excited with the optimal excitation
wavelength can emit blue, green, and orange light, respectively.

\vspace{-6pt}

\vspace{-6pt}

{The quantum yield $Q$ of B-CNDs, G-CNDs and O-CNDs can be evaluated
from the experimental data via \cite{Zhang17}
\begin{equation}\label{1}
Q=Q_{S}\cdot \frac{I_{S}}{I}\cdot \frac{A}{A_{S}}\cdot \frac{\eta}{\eta_{S}},
\end{equation}
where $Q_{S}$ is the quantum yield of the fluorescence for a standard sample for
reference. Under a fixed excitation wavelength at, e.g., 360 nm, 470 nm and 530 nm.
$I$ and $I_{S}$ are the integrated emission intensities of the CNDs sample and the standard
sample, respectively. $A$ and $A_{S}$ are respectively the absorbance of the prepared
sample and standard sample at the same excitation wavelength. $\eta$ and $\eta_{S}$ are
respectively the refractivity of the prepared sample and standard sample. The standard
sample of B-CNDs is quinine sulfate; the standard sample of G-CNDs and O-CNDs is
rhodamine. It is found that the fluorescent quantum yield of B-CNDs, G-CNDs and O-CNDs
is about 21.37$\%$, 16.12$\%$ and 9.11$\%$,~respectively.}

\begin{figure}[H]
\includegraphics[width=0.80\linewidth]{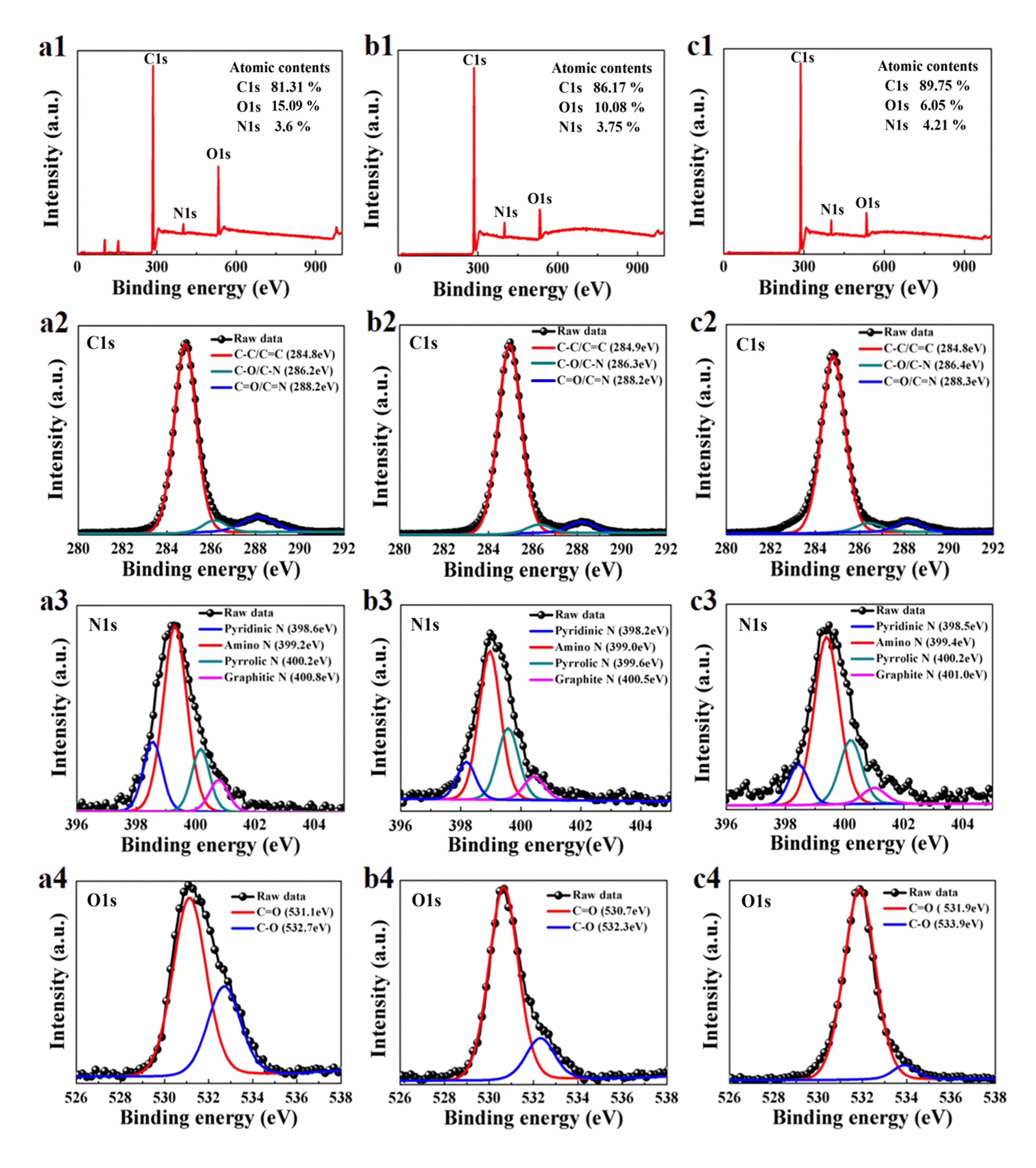}
\caption{The full-scan XPS spectra in (\textbf{a1}--\textbf{c1}), the high-resolution
spectra of C1s in (\textbf{a2}--\textbf{c2}), N1s in (\textbf{a3}--\textbf{c3}), and O1s in (\textbf{a4}--\textbf{c4}) for B-CNDs,
G-CNDs, and O-CNDs, respectively.}
\label{fig5}
\end{figure}

The XPS spectra and peak fitting diagrams of N-CNDs prepared by
citric acid and L-glutamic acid are shown in Figure \ref{fig5} with
the ratios of 1:1, 1:2, and 1:3. The XPS spectra have
three peaks at 285 eV, 399 eV, and 532 eV, which correspond to C1s, N1s,
and O1s peaks of the N-CNDs, respectively. The results reveal that
three N-CNDs samples are composed of C, N and O elements, and
the concentrations of these elements in B-CNDs, G-CNDs, and O-CNDs
were different. We can see that the nitrogen content increases with
increasing the ratio of citric acid to L-glutamic acid.
In Figure \ref{fig5}(a1--c1), it is worth noting that the fluorescence
wavelengths of N-CNDs show a red shift with the increase of the N doping
concentration. By comparing Figures \ref{fig3}a--c and
\ref{fig5}(a4--c4), it can be found that the fluorescence
intensity of N-CNDs is decreasing with decreasing the C-O groups on
the surface of N-CNDs.

However, the actual mechanism of fluorescence emission from CNDs is not
clear \cite{Zhu15}. It has been shown that the surface defects caused by
surface oxidation (C=O and C-O groups attached to CNDs) can be used as
the capture center of exciton, which could result in fluorescence emitting from
CNDs \cite{Bao15}. The elemental analysis in Figure \ref{fig5}(a1--c1)
shows that the contents of N in B-CNDs, G-CNDs, and O-CNDs samples
are 3.6\%, 3.75\%, 4.21\%, which increase regularly with increasing the ratio
between citric acid and L-glutamic acid. The N doping elements in CNDs
shown in Figure \ref{fig5}(a2--c3)
could also be the exciton capture centers which can change the surface state
of CNDs and lead to fluorescence similar to C=O and C-O
groups \cite{Guo16,Jiang15,Xu13}.

The deconvolution of high-resolution C1s XPS spectra in
Figure \ref{fig5}(a2--c2) reveals peaks at 284.8 eV, 286.4 eV,
and 288.2 eV, which correspond respectively to C-C/C=C, C-N/C-O,
and C=O/C=N bonding in N-CNDs \cite{Han17}. The stronger peak
of C-C/C=C indicates the better lattice structure of the $sp^2$
carbon (C-C/C=C) area. The high-resolution spectra of N1s for
N-CNDs in Figure \ref{fig5}(a3--c3)
contains Pyridinic N (398.5 eV),
Amino N (399.4 eV), Pyrrolic N (400.2 eV), and Graphite
N (401.0 eV) \cite{Yuan16,Ding15}. It is shown that a trace of
nitrogen atoms enters the carbon nucleus of N-CNDs and forms the PL
luminous center of N-CNDs through the hybrid of edge groups
with carbon nucleus \cite{Zhu15N}. The high-resolution spectra
of O1s in Figure \ref{fig5}(a4--c4) for N-CNDs contain
C=O (531.1 eV) and C-O (C-O-C/C-OH) (532.7 eV) \cite{Ding15,Zhang16}.
The quantity decreasing of C-O (C-O-C/C-OH) from a4 to c4 in
Figure \ref{fig5} is consistent with reducing the fluorescence
intensity in Figure \ref{fig3}a--c.
The intensity of PL emission changes with the peak's position
of C-O (C-O-C/C-OH).
The intensity of the PL emission
increases with increasing the peak height of C-O (C-OH), and with
reducing the peak height of O-C=O. This is in line with the
result we obtained previously \cite{Tang21}.
Therefore, it's reasonable to believe that
B/G/O-CNDs have various edge groups, such as C-OH, C-N, C=O, and C-H.
These edge groups can induce different kinds of surface states and
influence the intensity of PL spectra.
\vspace{-6pt}

\begin{figure}[H]
\includegraphics[width=0.8\linewidth]{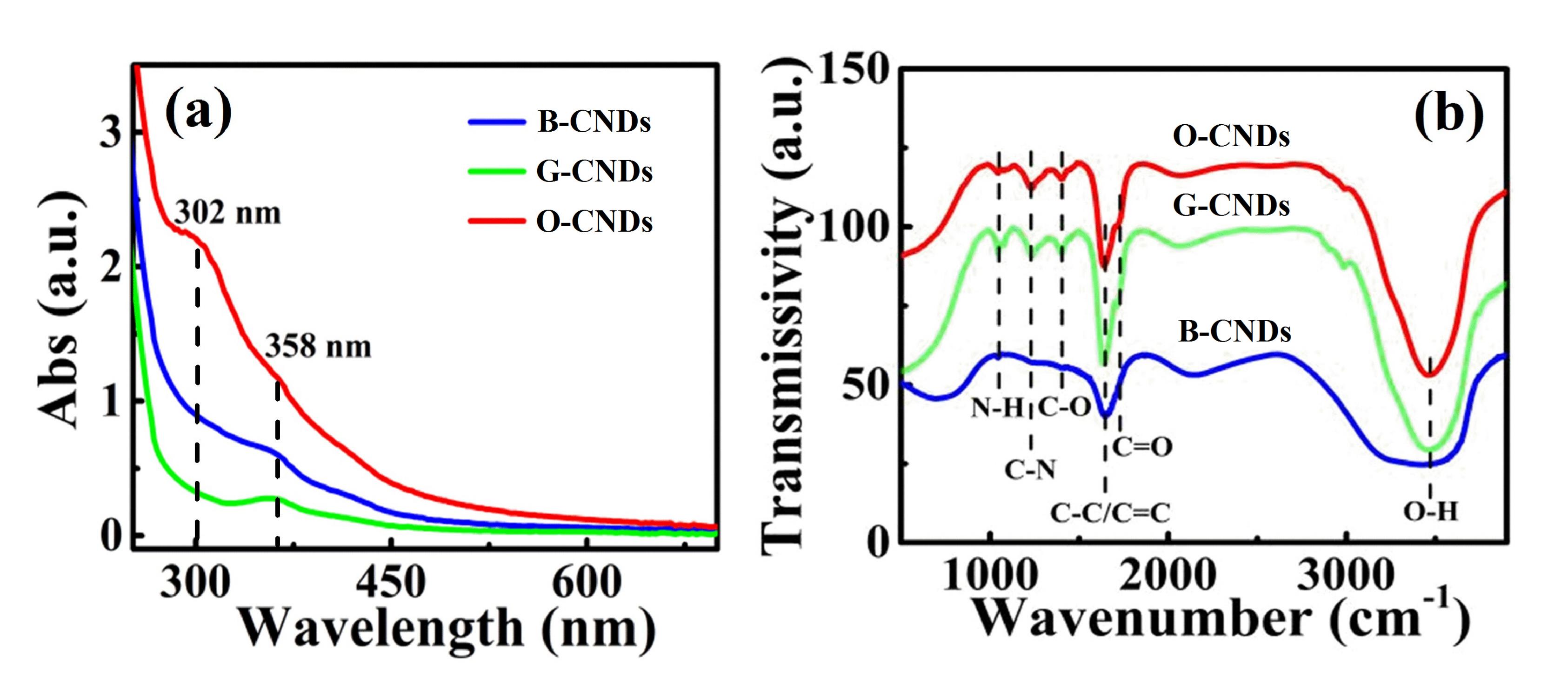}
\caption{UV-vis (\textbf{a}) and FT-IR (\textbf{b}) absorption spectra for N-CNDs.}
\label{fig6}
\end{figure}

The above results show that N-doping has effects on the light emission
of CNDs. As can be seen from Figure \ref{fig6}a, the absorption peak
of B-CNDs and G-CNDs is at 358 nm, and a clear absorption peak at 302 nm is
observed for O-CNDs. These absorption peaks are attributed by the functional
group (C=O) on the surface of N-CNDs \cite{Omer17}. These N-CNDs have
different surface states, and the light emissions may originate from
the radiation recombination of the excited electrons \cite{Hu15}.
Thus, the PL emission properties of N-CNDs is mainly attributed to
the N-doping effect. The surface groups of N-CNDs prepared under microwave
conditions were analyzed by FT-IR \cite{Tucureanu16} and shown in
Figure \ref{fig6}b. It can be shown that the stretching vibrations
of O-H are at 3400--3200 cm$^{-1}$, which indicates that N-CNDs are
rich in hydroxyl groups. The characteristic absorption frequency
of C=O (1600--1900~cm$^{-1}$) near 1646 cm$^{-1}$ overlaps with that
of C=C/C-C (1500--1675 cm$^{-1}$). We can understand that there are carboxy
groups around the carbon spot. The vibration absorption peak of
N-C$_3$ is at 2063 cm$^{-1}$. The absorption peak of C-O bond
stretching vibration is at 1401 cm$^{-1}$, and this is the evidence
of the polymerization between citric acid or L-glutamic acid.
The bending vibration absorption peak of C-N is at 1231 cm$^{-1}$
and the absorption peak of C-O-C is at 1045 cm$^{-1}$.
The results in Figure \ref{fig6} imply that the particles are
surrounded by functional groups, such as hydroxyl groups and carboxy
groups. These functional groups not only improve the water solubility
and biocompatibility of N-CNDs, but also contribute to the surface
modification of N-CNDs. These are in good agreement with the results
obtained from XPS spectra.

\begin{figure}[H]
\includegraphics[width=0.43\linewidth]{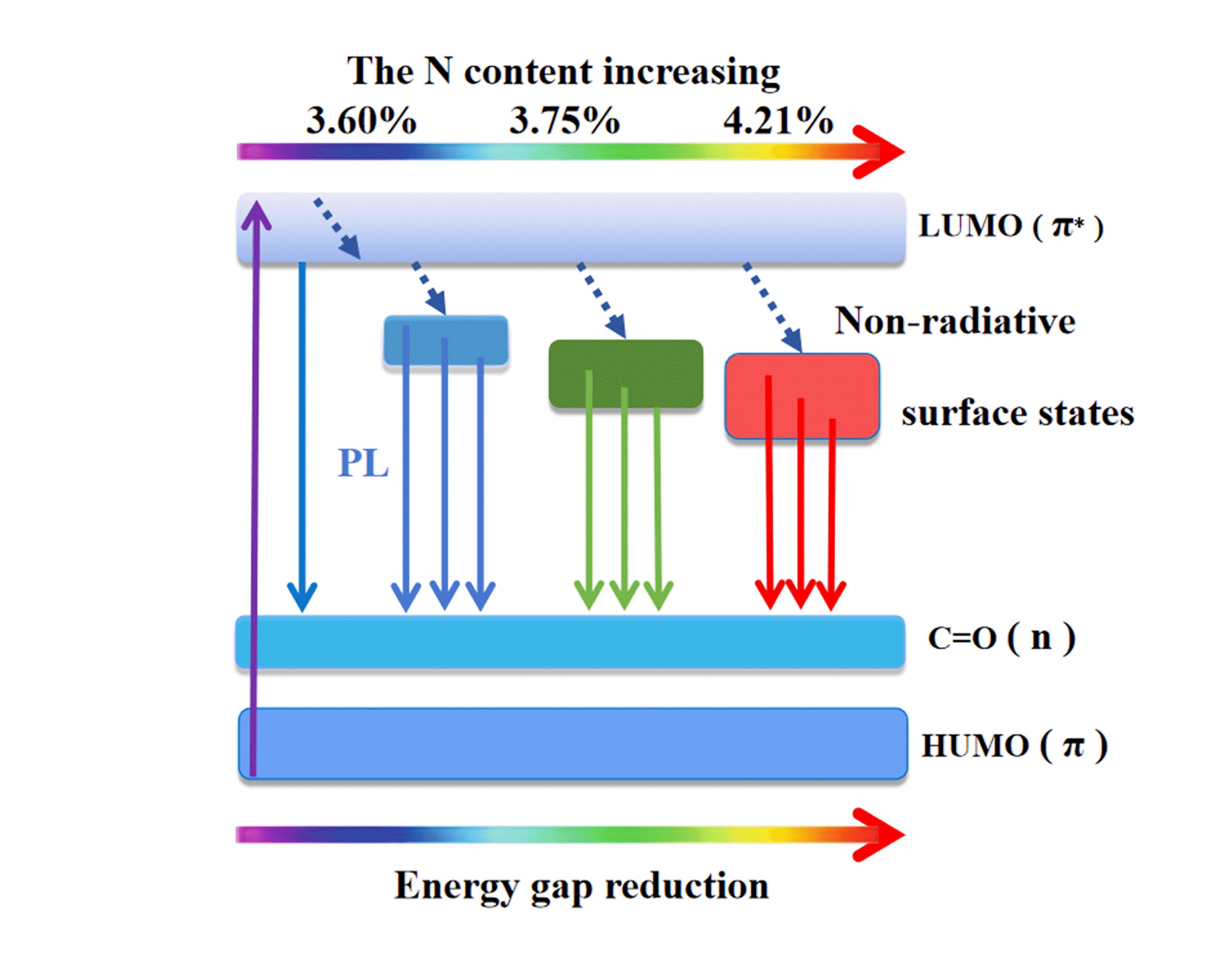}
\caption{The sketch diagram of color tuning mechanism in N-CNDs.}
\label{fig7}
\end{figure}

Figure \ref{fig7} shows the color tuning mechanism of N-CNDs. With
increasing the N doping concentration of N-CNDs, more surface states are
introduced, and the lowest energy level of surface states
decreases \cite{Purbia16}. The fluorescence wavelength of N-CNDs would
have red-shifts. As a result, the fluorescence wavelengths of N-CNDs
can be tuned through controlling N-doping concentrations of N-CNDs.
Multitudinous surface states on N-CNDs result in a wide range of energy
bands, which correspond to broad absorption bandwidths and the
excitation wavelength-dependent emission spectra.
From the viewpoint of physics, CNDs have strictly discrete electron
energy levels, and the filling distribution of electrons determines the
molecular states. The molecular states are divided into singlet and triplet
states. At room temperature, most of the bonding electrons in the molecules
of fluorescent substances are in the ground state. Under the action of optical
shining, some of the bonding electrons in the ground state move
to a higher electron excited state, resulting in the phenomenon of molecular
absorption in spectrophotometry. The electrons at the excited state energy level
are in the non-ionized state. The electrons in the first excited state can
transition to the second excited state, through non radiative transitions such
as internal conversion and vibrational relaxation. In the process of decaying
back to each vibrational level in the ground state, they release energy or photons
and produce fluorescence.

{As can be seen, the surface of N-CNDs prepared in the work contains a large number of
functional groups, which can make N-CNDs show a strong affinity for biomass.
Therefore, N-CNDs can be dispersed and adsorbed on biomass materials, such as paper
plates and clothing walls, which can emit bright fluorescence under UV light
excitation. Therefore, N-CNDs would have potential application value in interior decoration,
anti-counterfeiting trademark, and other fields.}

\section{Conclusions}

In conclusion, we have prepared N-CNDs by microwave method with citric acid
and L-glutamic acid as carbon source, and deionized water as a dispersing agent.
The optical properties of different N-doping CNDs have been investigated with
different ratios between citric acid and L-glutamic acid. The results indicate
that many functional groups, such as C-O, C=O, C-O-C, C-N attached
to the surface/edges of carbon nucleus of N-CNDs, exist. The prepared N-CNDs with
the ratios of the raw materials at 1:1, 1:2, and 1:3 can emit blue (445 nm),
green (546 nm), and orange (617 nm) fluorescence, respectively.
Different N concentrations doped into N-CNDs can lead to different surface/edge states, and
the wavelength of the PL emission from N-CNDs can be tuned via controlling the N
concentration in CNDs. The PL intensity of N-CNDs is associated with the content of C-O
groups on the surface/edges of N-CNDs. The wavelength of the PL emission from
N-CNDs shows a red-shift, with increasing of the N concentration in N-CNDs.
The interesting and important findings from this study can help us to gain
a better understanding of the microscopic mechanism for achieving multi-colour
fluorescent CNDs.

\vspace{6pt}

\authorcontributions{W.L. and J.T. prepared the samples and carried out the
experimental measurements. Y.L., H.B. and W.Z. participated in
the sample preparation and the analyses of the experimental results.
J.Z. proposed the research work, designed the experiments and prepared
the manuscript. Y.X. and W.X. participated in the analyses of the experimental
results and the preparation of the manuscript. All authors read and approved
the final manuscript.}

\funding{This work was supported by the National Natural Science Foundation
of China (grant Nos. 12064049, 11664044, U1930116, U1832153,
U206720039, 12004331, 11847054), the Department
of Science and Technology of Yunnan Province (grant Nos. 202004AP080053,
2019FD134), Shenzhen Science and Technology Program
(KQTD20190929173954826).}

\conflictsofinterest{The authors declare no conflict of interest.}

\reftitle{References}

\end{document}